\documentclass[aps,twocolumn,amsmath,amssymb, superscriptaddress]{revtex4}
\usepackage[pdftex]{graphicx}
 \usepackage{amsmath}
\usepackage[T1]{fontenc}
\usepackage{amssymb}
\usepackage{amsfonts}
\usepackage{flushend}
 \usepackage{amsmath} 
\usepackage[breaklinks=true,colorlinks=true,linkcolor=blue,urlcolor=blue,citecolor=blue]{hyperref}
\usepackage{lipsum}
\usepackage{amsfonts} 
\usepackage{amssymb, mathrsfs}
\usepackage{braket}
\usepackage{graphicx} 
\usepackage{subfigure}
\usepackage{bbm}
\def\beq{\begin{equation}}
\def\eeq{\end{equation}}
\def\bsp{\begin{split}}
\def\esp{\end{split}}
\def\bea{\begin{eqnarray}}
\def\eea{\end{eqnarray}}
\def\ba{\begin{array}}
\def\ea{\end{array}}

\def\lb{\left(}
\def\rb{\right)}

\def\l.{\left.}
\def\r.{\right.}

\def\ra{\rangle}
\def\la{\langle}

\def\bo{\bold{k}}

\begin{document}

\title{Chirality-induced magnon transport in AA-stacked bilayer honeycomb  magnets}
\author{S. A. Owerre}
\affiliation{Perimeter Institute for Theoretical Physics, 31 Caroline St. N., Waterloo, Ontario N2L 2Y5, Canada.}
\email{sowerre@perimeterinstitute.ca}
\affiliation{African Institute for Mathematical Sciences, 6 Melrose Road, Muizenberg, Cape Town 7945, South Africa.}

\begin{abstract}
   In this Letter, we study the magnetic transport in AA-stacked bilayer honeycomb chiral magnets coupled either ferromagnetically or antiferromagnetically. For both couplings, we observe chirality-induced gaps,  chiral protected edge states,  magnon Hall and magnon spin Nernst effects of magnetic spin excitations.  For ferromagnetically coupled layers, thermal Hall and spin Nernst conductivities do not change sign as function of magnetic field or temperature similar to single-layer honeycomb ferromagnetic insulator. In contrast, for antiferromagnetically coupled layers, we observe a sign change in the thermal Hall and spin Nernst conductivities as the magnetic field is reversed.  We discuss possible experimental accessible honeycomb bilayer quantum materials in which these effects can be observed. \\\\
   Online supplementary data available from \href{http://stacks.iop.org/JPhysCM/28/47LT02/mmedia}{stacks.iop.org/JPhysCM/28/47LT02/mmedia}
\end{abstract}
\maketitle

\section{Introduction}
  The subject of chirality is ubiquitous in many branches of physics. It plays a prominent role in the understanding of many physical systems, in particular magnetic systems. Recently, the influence of spin chirality in magnon excitations of ordered quantum magnetic materials has attracted considerable attention. Chirality-induced topological magnetic systems  exhibit novel properties \cite{alex0,alex1, alex1a, zhh, alex2, alex4, alex4h, mak5,shin,shin1} similar to topological insulators in electronic systems \cite{yu6,yu7, fdm}. In these systems,  the propagation of collective excitations (magnons) is similar to spin-orbit coupling induced propagation  in non-interacting electronic systems. However, non-interacting magnons can  propagate without dissipation as they are uncharged quasiparticles. This property  is crucial for future dissipationless magnon transport \cite{kru,kru1,kru2, alex1}. In magnetic structures with corner sharing triangles, chirality is induced by the Dzyaloshinsky-Moriya interaction (DMI)  \cite{dm} and spin scalar chirality is defined as $\chi_{ijk}={\bf S}_i\cdot ({\bf S}_j\times {\bf S}_k)$, where $(i,j,k)$ label sites on the triangular plaquettes of the lattice and ${\bf S}_i$ is the spin at site $i$. 

Recently, chirality-induced transports have been observed in two-dimensional (2D) single-layer kagome magnet Cu(1-3, bdc) \cite{alex6a, alex6}.  Also  in a number of 3D single-layer chiral ferromagnetic pyrochlores such as Lu$_2$V$_2$O$_7$, Ho$_2$V$_2$O$_7$, and In$_2$Mn$_2$O$_7$ \cite{alex1,alex1a}. For ferromagnetic pyrochlores, thermal Hall conductivity changes sign upon reversing the direction of magnetic field, whereas for kagome magnet thermal Hall conductivity changes sign as function of magnetic field or temperature. There has been a great interest in these chirality-induced topologically protected magnetic systems \cite{xc,sol,sol1,kkim}. Recently, chirality-induced magnetic bilayer-skyrmion lattice has been shown to exhibit  novel properties completely different from single-layer skyrmions \cite{mak5}. However, transport properties of chirality-induced magnetic bilayer magnons have not been addressed in many chiral magnetic systems.  

In this Letter, we study the chirality-induced magnetic transport in AA-stacked bilayer honeycomb chiral magnet coupled either ferromagnetically or antiferromagnetically.  For ferromagnetically coupled layers, we compute the thermal Hall conductivity $\kappa_{xy}$,  which shows a similar trend to that of  single-layer honeycomb ferromagnetic insulator \cite{sol,sol1} with no sign change. In contrast, for antiferromagnetically coupled layers, we observe a sign change in the thermal Hall conductivity and spin Nernst conductivity \cite{alex7} as the magnetic field is reversed. We show that fully antiferromagnetically coupled  bilayer  with ordered N\'eel state  exhibits similar properties to that  of antiferromagnetically coupled bilayer ferromagnets.  These results suggest an experimental search for chirality-induced bilayer honeycomb chiral magnets. In fact, many experimental realizations of bilayer honeycomb-lattice systems  have been reported. They include magnetic compounds such as CrBr$_3$ \cite{dav0,dav} which is a spin-$1/2$ bilayer honeycomb ferromagnetic materials. Na$_3$Cu$_2$SbO$_6$ \cite{aat1} and $\beta$-Cu$_2$V$_2$O$_7$ \cite{aat} are spin-$1/2$ Heisenberg antiferromagnetic materials, in each of which the $S = 1/2$ Cu$^{2+}$ ions are situated on the sites of weakly coupled honeycomb-lattice layers. Besides, the iridates A$_2$IrO$_3$ (A= Na, Li) \cite{aat0,aat00} have  magnetically ordered Mott phases in which the Ir$^{4+}$ ions form effective $S=1/2$ moments arrayed on weakly-coupled honeycomb-lattice layers.  In these honeycomb-lattice materials, spin chirality or DMI can be induced   using many experimental growth techniques and our results can be confirmed directly.

\section{Bilayer ferromagnetic insulator}
\begin{figure*}[!]
\centering
  \subfigure[\label{lattice1}]{\includegraphics[width=.35\linewidth]{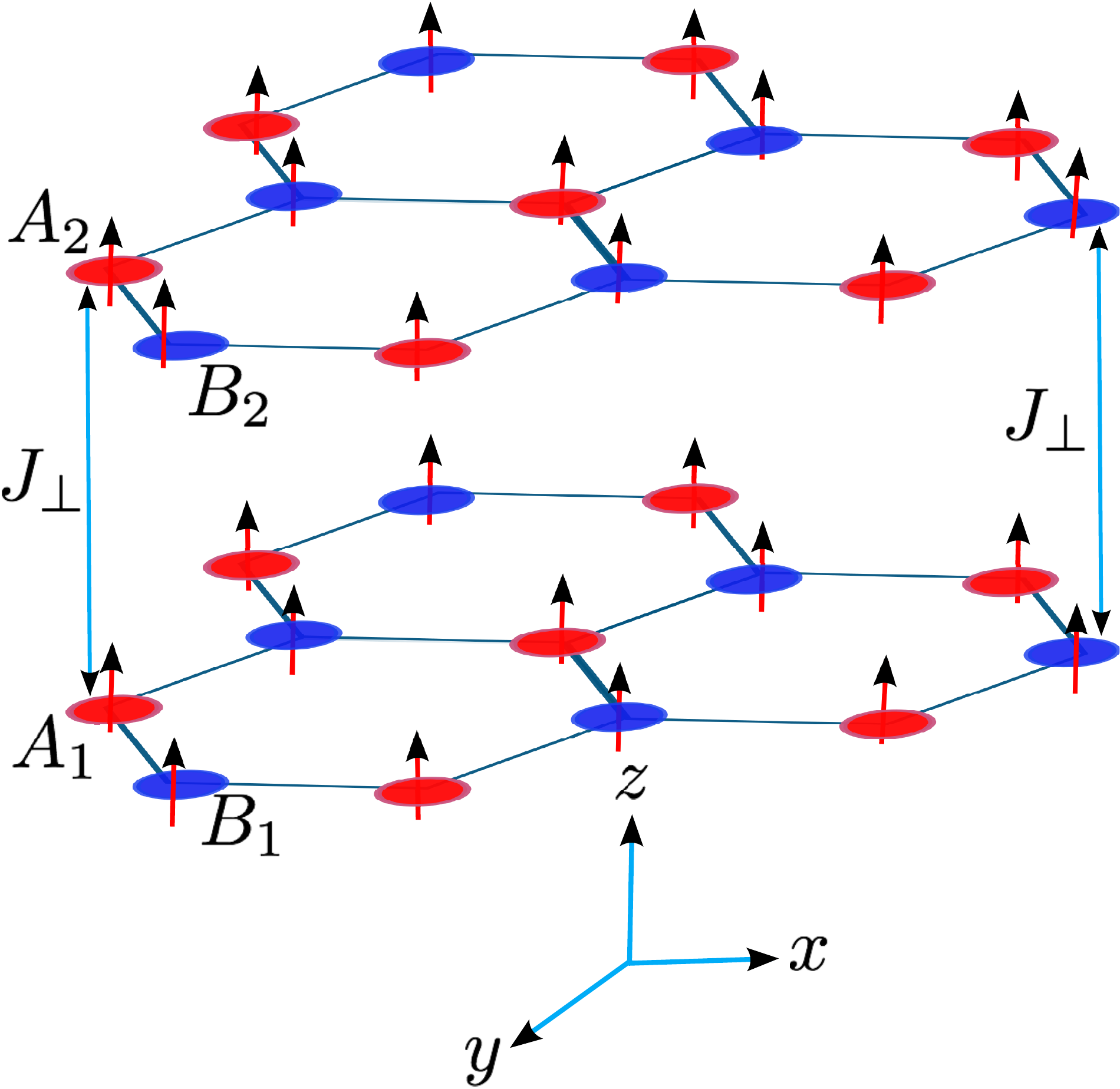}}
  \quad\quad
   \subfigure[\label{lattice2}]{\includegraphics[width=.35\linewidth]{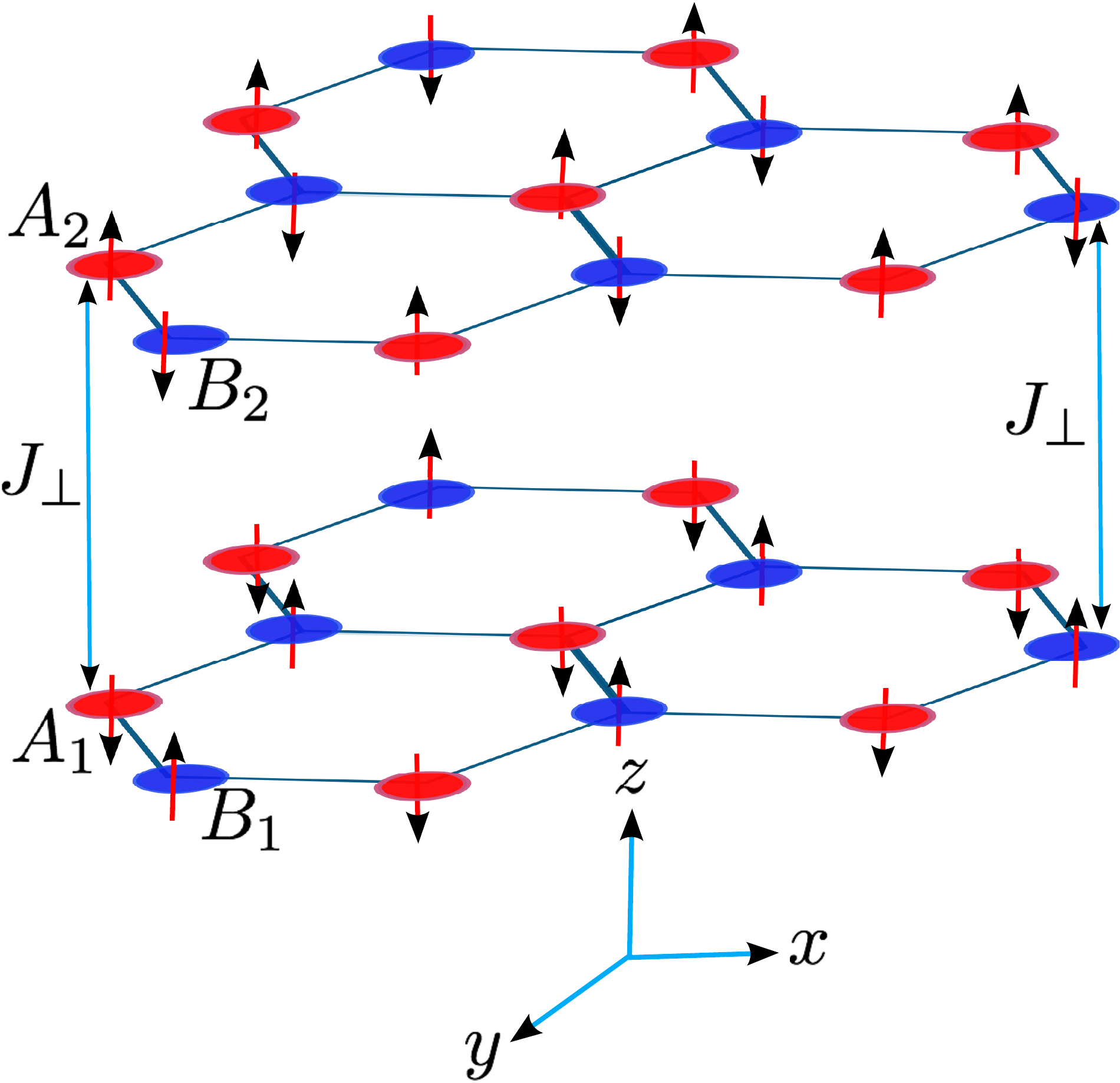}}
\caption{Color online. $(a)$  Ferromagnetically coupled AA-stacked  bilayer honeycomb magnets. For antiferromagnetically coupled system, the spins on the upper or layer are pointing downwards. $(b)$  AA-stacked honeycomb-lattice bilayer  N\'eel antiferromagnet.}
\end{figure*}

 The Hamiltonian for AA-stacked bilayer honeycomb chiral magnet shown in Fig.~\ref{lattice1} is given by
\begin{align}
H=H_{FM}^\tau +H_{DM}^\tau +H_{ext.}^\tau+ H_{inter}
\end{align}
where
\begin{align}
H_{FM}^\tau &=-J\sum_{\la i, j\ra}{\bf S}_{i}^\tau\cdot{\bf S}_{j}^\tau-J^\prime\sum_{\la\la i, j\ra\ra}{\bf S}_{i}^\tau\cdot{\bf S}_{j}^\tau \label{h2}\\ H_{DM}^\tau&=\sum_{\la \la i,j\ra\ra} {\bf D}_{ij}\cdot{\bf S}_{i}^\tau\times{\bf S}_{j}^\tau;~H_{ext.}^\tau =-h\sum_{i}S_{i,z}^\tau,\\H_{inter.}&=-J_{\perp}\sum_{ i}{\bf S}_{i}^T\cdot{\bf S}_{i}^B,\label{h}
\end{align}
where $\tau$ denotes the top ($T$) and bottom ($B$) layers. ${\bf S}_{i}$ is the spin moment at site $i$, $J>0$ is a nearest-neighbour (NN) ferromagnetic interaction on each layer, $J^\prime>0$ is a next-nearest-neighbour (NNN) ferromagnetic interaction on each layer, and $\bold{D}_{ij}$ is the DMI vector between site $i$ and $j$, allowed by the NNN triangular plaquettes on the honeycomb lattice, where  ${\bf D}_{ij}=\nu_{ij}{\bf D}$, and $\nu_{ij}=\pm 1$.   The Zeeman magnetic field is $h$ in units of $g\mu_B$. The interaction $J_{\perp}>0$ represents the ferromagnetic or antiferromagnetic ($J_{\perp}<0$) interlayer coupling. 


\section{Magnon bands}
 \subsection{Ferromagnetic interlayer coupling}
\begin{figure}[ht]
\centering
\includegraphics[width=3in]{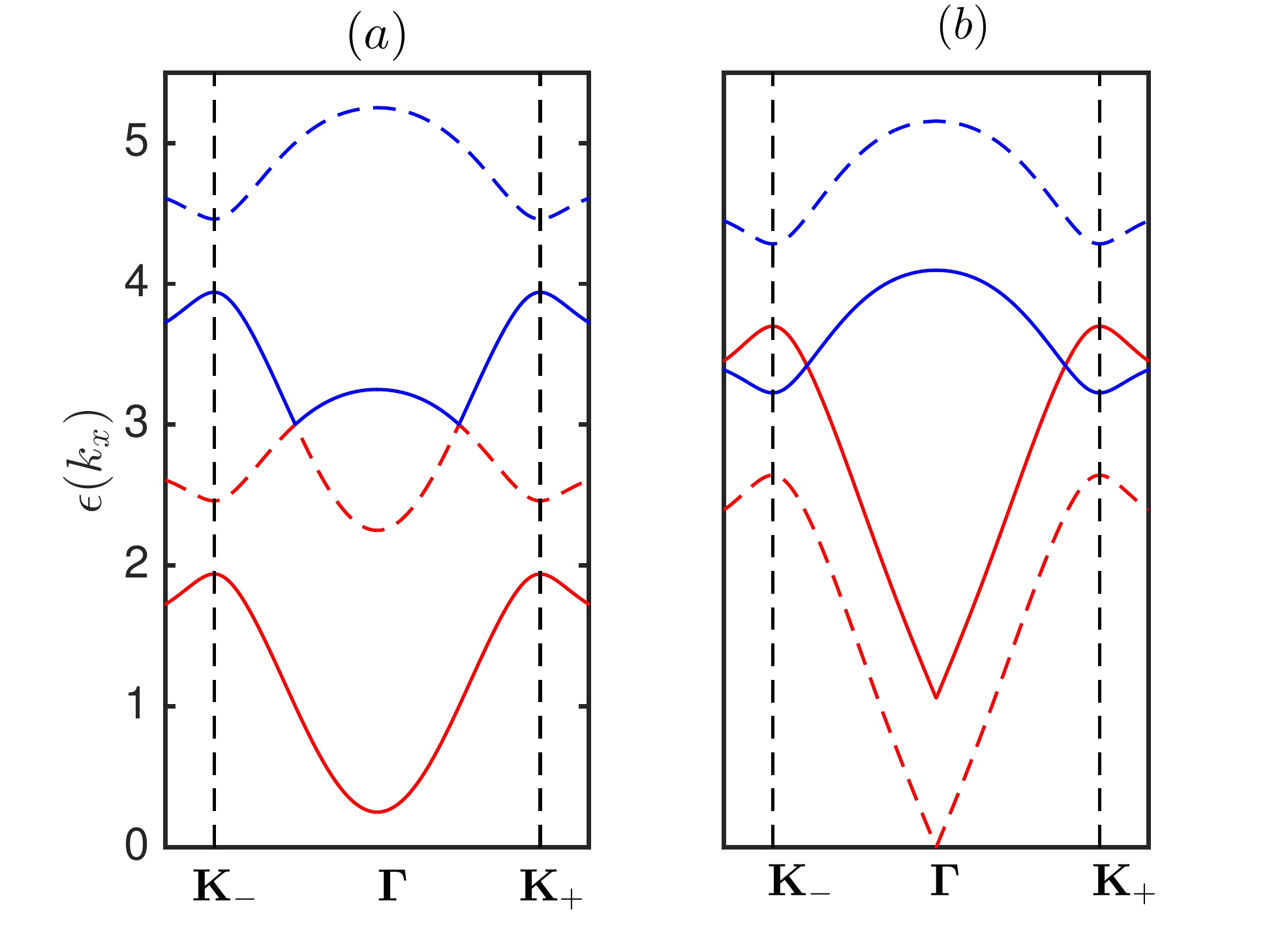}
\caption{Color online. Magnon band structures of spin-$1/2$  bilayer honeycomb chiral magnet for  $h=v_s=0.5$, $v_D=v_s^\prime=0.05,~v_\perp=1,~\phi=\pi/4$. $(a)$~Ferromagnetic coupling. $(b)$~ Antiferromagnetic coupling.}
\label{bands}
\end{figure}

For ferromagnetic interlayer coupling $J_\perp >0$,  the Fourier space Hamiltonian of the Holstein-Primakoff \cite{HP} boson operators is given by $ H=\sum_{\bold k}\psi^\dagger_{\bold k}\cdot \mathcal{H}(\bold k)\cdot\psi_{\bold k}$ where $\psi^\dagger_{\bold k}= (a_{\bold{k}A_1}^{\dagger},\thinspace a_{\bold{k} B_1}^{\dagger},a_{\bold{k} A_2}^{\dagger},\thinspace a_{\bold{k} B_2}^{\dagger})$, and
 \begin{align}
\mathcal{H}_{FM}(\bold k)&=\epsilon_{a}\tau_0\otimes\sigma_0+ \tau_0\otimes\sigma_z m_{\bo\phi}-v_s\tau_0\otimes(f_\bo\sigma_+  + f_\bo^*\sigma_{-})\nonumber\\&-v_\perp\tau_x\otimes\sigma_0,
\label{honn}
\end{align}
where  $\boldsymbol{\sigma}$ and $\boldsymbol{\tau}$ are triplet  pseudo-spin Pauli matrices for the sublattice and layer degrees of freedom respectively, $\tau_0$ and $\sigma_0$ are identity matrices in each space, and $\sigma_\pm=(\sigma_x\pm i\sigma_y)/2$;  $v_0=h+zv_s+z^\prime v_s^\prime$,  $v_s(v_s^\prime)(v_D)(v_\perp)= JS(J^\prime S)(DS)(J_\perp S)$, $v_t=\sqrt{v_s^{\prime 2} +v_D^2}$, $z(z^\prime)=3(6)$, and $\epsilon_{a}=v_0+v_\perp-2v_t\sum_{\mu} \cos(\bo\cdot\bold{a}_\mu) \cos\phi$. Here, $f_\bo= e^{ik_ya/2}\lb 2\cos\sqrt{3}k_xa/2+e^{-3ik_ya/2}\rb,$ and  $m_{\bo\phi}= 2v_t\sum_{\mu} \sin(\bo\cdot\bold{a}_\mu) \sin\phi$, where $\bold a_1=\sqrt{3}\hat x;~ \bold a_2=(-\sqrt{3}\hat x, {3}\hat y)/2~ \bold a_3=-(\sqrt{3}\hat x, {3}\hat y)/2$.  We have assumed a DMI along the $z$-axis.   The phase factor  $\phi=\arctan(D/J^\prime)$ is a magnetic flux generated by the DMI on the NNN triangular plaquettes. The eigenvalues are given by
\begin{align}
\epsilon_{\alpha \pm}({\bf k})=\epsilon_a +(-1)^\alpha v_\perp \pm\sqrt{m_{\bo\phi}^2+|v_sf_\bo|^2},
\label{fm}
\end{align}
where $\alpha=1,2$ is the layer index.    For $v_\perp=0$, the Hamiltonian decouples to two single layers \cite{sol}. The magnon band is shown  in  Fig.~\ref{bands}(a).       At the Dirac points ${\bf K}_\pm= (\pm 4\pi/3\sqrt{3}a, 0)$, the eigenvalues reduces to  $\epsilon_{\alpha\pm}({\bf K}_\pm)=\epsilon_{a}+(-1)^\alpha v_\perp\pm   |m_\phi|,$ where $m_\phi=3\sqrt{3}v_t\sin\phi$.  This is similar to AA-stacked spin-orbit-coupled bilayer graphene \cite{mak55, mak66}. 
  \subsection{Antiferromagnetic interlayer coupling}

We now consider  antiferromagnetically coupled layers, with the spins on the upper or lower layer pointing downwards, and the  interlayer coupling $J_\perp<0$.   The top and bottom layers are still ferromagnetic insulators described by $H_{FM}^\tau$ and $H_{DM}^\tau$. To study the magnetic excitations, we perform  a $\pi$-rotation about  the $S_x$-axis on the top layer, \bea S_{i}^x&\to S_{i}^x,~ S_{i}^y\to -S_{i}^y,~S_{i}^z\to -S_{i}^z\label{ro3}.\eea   This rotation keeps the upper ferromagnetic layer invariant but points the spins in  the new $z$-direction, and changes the sign of  $H_{ext}^\tau$ and $H_{DM}^\tau$ on the top layer.  The Fourier space  Hamiltonian is $ H=\frac{1}{2}\sum_{\bold k}\Psi^\dagger_{\bold k} \mathcal{H}_{AFM}(\bold k)\Psi_{\bold k}+\text{const.}\label{ham1}$, where  $\Psi_{\bold k}=(\psi_\bo^\dagger,~\psi_{-\bo})$,  and
\begin{align}
\mathcal{H}_{AFM}(\bold k)=\left(
\begin{array}{cc}
\mathcal A(\bo)& \mathcal B\\
\mathcal B& \mathcal A^*(-\bo)
\end{array}
\right).
\label{atn}
\end{align}
The matrices $\mathcal A(\bo)$ and $\mathcal B$ are given by
\begin{align}
&\mathcal A(\bo)=\epsilon_{0}\tau_0\otimes\sigma_0+ \tau_z\otimes\sigma_z m_{\bo\phi}-v_s\tau_0\otimes(f_\bo\sigma_+ + f_\bo^*\sigma_-)\nonumber\\&+h\tau_z\otimes\sigma_0;~\mathcal B= |v_\perp|\tau_x\otimes\sigma_0,
\label{mata}
\end{align}
where $\epsilon_{0}=zv_s+z^\prime v_s^\prime+ |v_\perp|-2v_t\sum_{\mu} \cos(\bo\cdot\bold{a}_\mu) \cos\phi$. Note that $\mathcal A(\bold k)\neq \mathcal A^*(-\bold k)$ due to the DMI.  The Hamiltonian is diagonalized by generalized Bogoliubov transformation (see Supplemental material: \href{http://stacks.iop.org/JPhysCM/28/47LT02/mmedia}{stacks.iop.org/JPhysCM/28/47LT02/mmedia}).  The positive  eigenvalues are given by
\begin{align}
\epsilon_{\alpha,\pm }({\bf k})= (-1)^\alpha h+\sqrt{\bigg[\epsilon_0\pm \sqrt{m_{\bo\phi}^2+|v_sf_\bo|^2}\bigg]^2-v_\perp^2}.
\label{afm}
\end{align}
 For $h=0$, the energy bands are doubly degenerate --- one of the major differences between ferromagnetically and antiferromagnetically coupled layers. Also notice that  antiferromagnetically coupled layers have a linear dispersion near ${\bf \Gamma}$ (see Fig.~\ref{bands}(b) ) as opposed to a quadratic dispersion in the ferromagnetic case.    For $v_\perp=0$, Eq.~\ref{afm} decouples and reduces to Eq.~\ref{fm} with opposite magnetic field on each layer.

\subsection{Bilayer  antiferromagnetic insulator}

 In the fully antiferromagnetic case, each layer is modeled by the Heisenberg antiferromagnet, with  $J,J^\prime, J_\perp<0$. Due the $J^\prime$ term, the Heisenberg antiferromagnet is frustrated as opposed to the ferromagnetic counterpart.  With zero DMI  $H_{DM}^\tau =0$,  the system is considered to describe bilayer honeycomb antiferromagnetic material Bi$_3$Mn$_4$O$_{12}$(NO$_3$) \cite{mak0, mak1, mak2, mak3, mak4}. The ground state phase diagram of this model has been studied extensively \cite{mak0, mak1, mak2, mak3, mak4}. It consists of an ordered N\'eel state for $J^\prime/J<1/6$ and a nonmagnetic state for  $J^\prime/J>1/6$ \cite{mak0}. For large values of $J_\perp$, the ground state is an interlayer valence-bond crystal in which the  spins from both layers form dimers \cite{mak3}. 

We are interested in the topological effects of the  ordered N\'eel state for $J^\prime/J<1/6$ shown in Fig.~\ref{lattice2}. Such N\'eel state order exists in the bilayer honeycomb iridates  A$_2$IrO$_3$ (A= Na, Li) \cite{aat0,aat00}. In this phase, the band structure in the absence of the chiral DMI exhibits Dirac points at ${\bf K}_\pm= (\pm 4\pi/3\sqrt{3}a, 0)$ \cite{mak1, mak2,mak3, mak4}.   A nearest-neighbour DMI does not introduce chirality and an external magnetic field introduces canting  up to the saturated field when fully polarized ferromagnetic states are recovered. These terms  do not open a gap at ${\bf K}_\pm$.  As in the ferromagnetic case, chirality is introduced  by a next-nearest-neighbour DMI. As we now show, this is very similar to antiferromagnetically coupled  bilayer ferromagnets studied above at zero magnetic field.  The only difference is that the NNN coupling is restricted to $J^\prime/J<1/6$.

We begin by performing the $\pi$-rotation describe above on sublattice $A_1$ and $B_2$ such that the  spins point along the new rotated $z$-axis. The SU(2)-invariant NN and NNN interactions on each layer are invariant under this rotation, but the U(1)-invariant out-of-plane DMI changes sign as in the previous case. In the bosonic representation, the Hamiltonian has the form as Eq.~\ref{atn} with  
\begin{figure}[ht]
\centering
\includegraphics[width=3in]{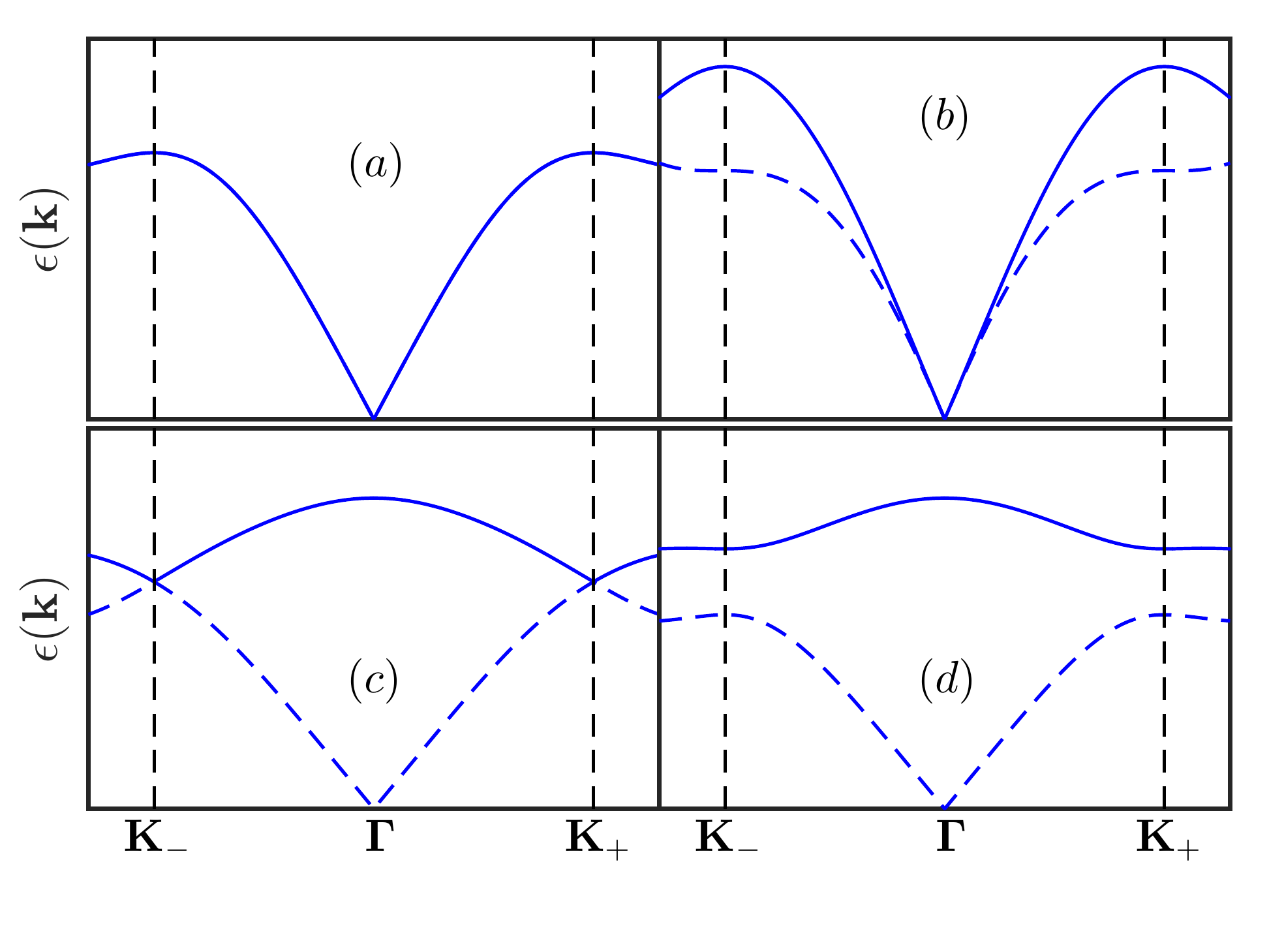}
\caption{Color online. Magnon bulk bands of the spin-$1/2$ AA-stacked bilayer N\'eel antiferromagnet along $k_y=0$ at $h=0$. The parameters are $v_s=0.5$ $(a)$. $v_D=v_\perp=0$; $v_s^\prime=0.05$. $(b)$. $v_\perp=v_s^\prime=0.0,~v_D=0.05,~\phi=\pi/2$. $(c)$. $v_\perp=v_s^\prime=0.05,~v_D=0.0,~\phi=0$; $(d)$. $v_D=v_s^\prime=0.05,~v_\perp=0.5,~\phi=\pi/4$.}
\label{bandsA}
\end{figure}
\begin{figure*}[ht]
\centering
  \subfigure[\label{e0}]{\includegraphics[width=.47\linewidth]{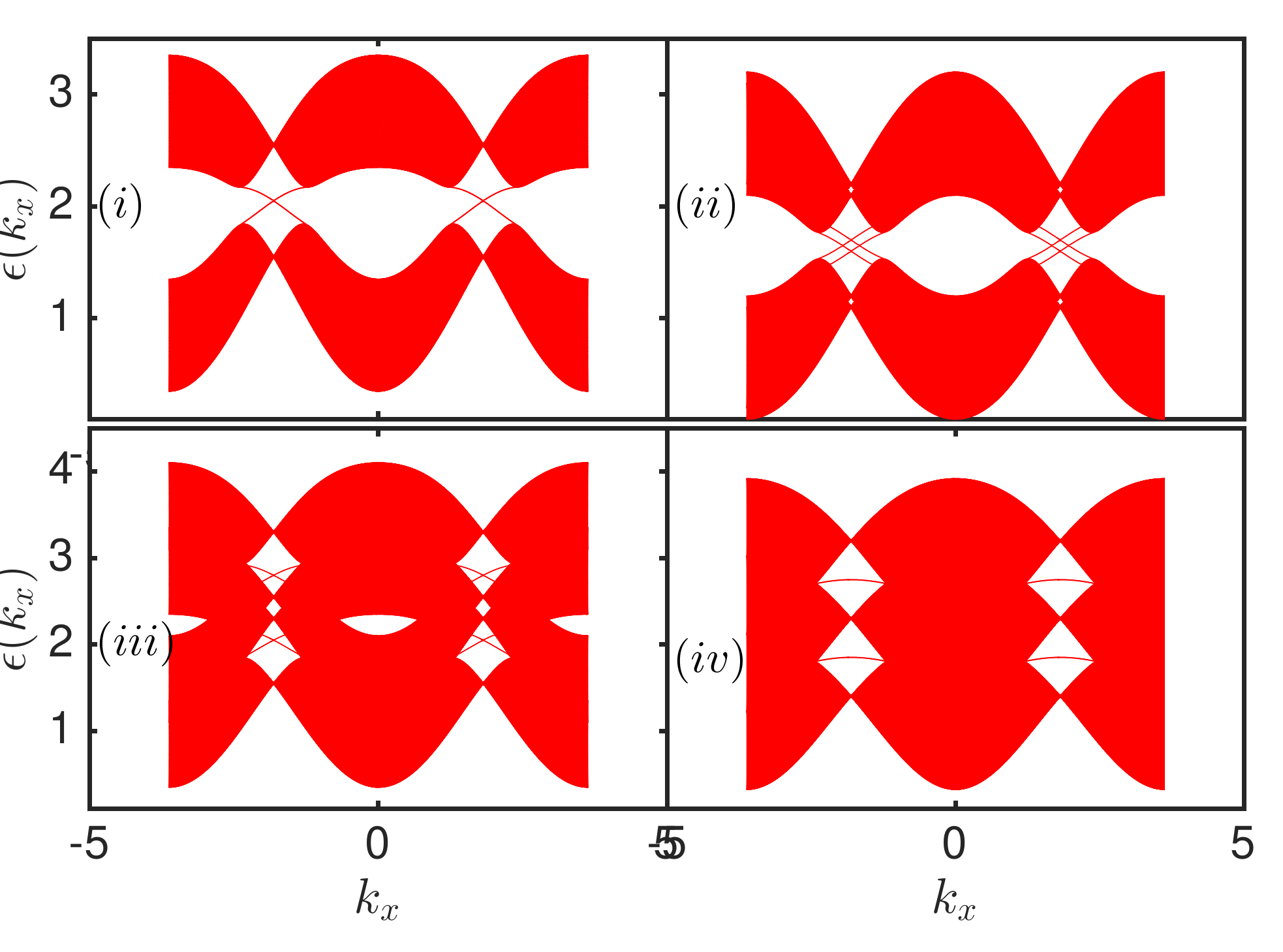}}
   \subfigure[\label{ee0}]{\includegraphics[width=.47\linewidth]{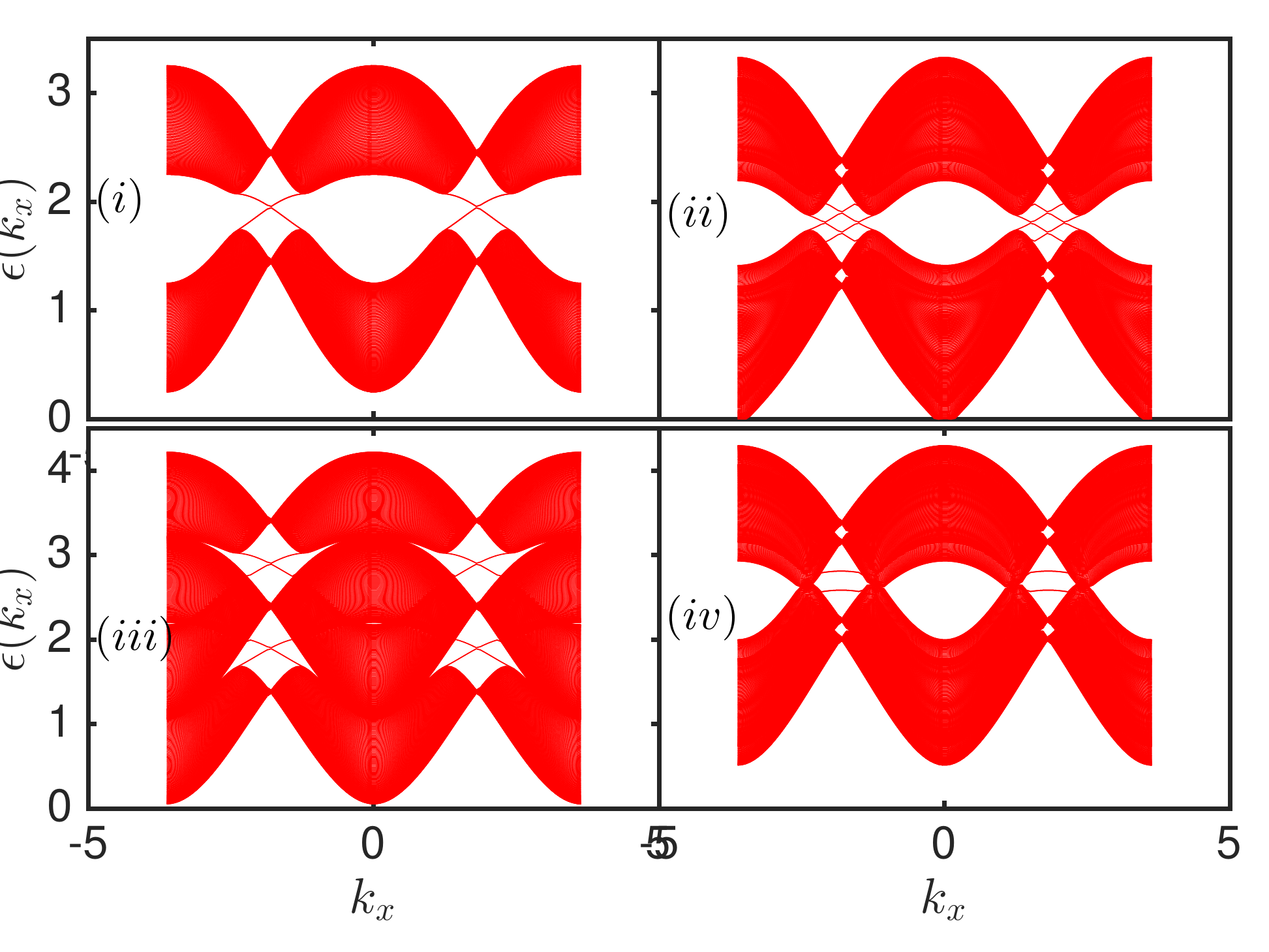}}
\caption{Color online.  Chiral edge states of  AA-stacked bilayer honeycomb chiral magnets with $v_s=0.5$. $(a)$ Ferromagnetically coupled  layers $(i)~v_D=0.1,~v_s^\prime=0.05, ~h=0.1, ~v_\perp=0$. $(ii)~v_D=0.1,~v_s^\prime=0.0, ~h=0.1, ~v_\perp=0.05$. $(iii)~v_D=0.1,~v_s^\prime=0.05, ~h=0.1, ~v_\perp=0.375$. $(iv)~v_D=0.0,~v_s^\prime=0.05, ~h=0.1, ~v_\perp=0.5$.   $(b)$ Antiferromagnetically coupled  layers $(i)~v_D=0.1,~v_s^\prime=0.05,~h=v_\perp=0$. $(ii)~v_D=0.1,~v_s^\prime=0.0, ~h=0.1, ~v_\perp=0.5$. $(iii)~v_D=0.1,~v_s^\prime=0.05, ~h=0.5, ~v_\perp=0.5$. $(iv)~v_D=0.0,~v_s^\prime=0.05, ~h=0.1, ~v_\perp=0.5$.}
\end{figure*}
\begin{figure}[ht]
\centering
  \subfigure[\label{e1}]{\includegraphics[width=.45\linewidth]{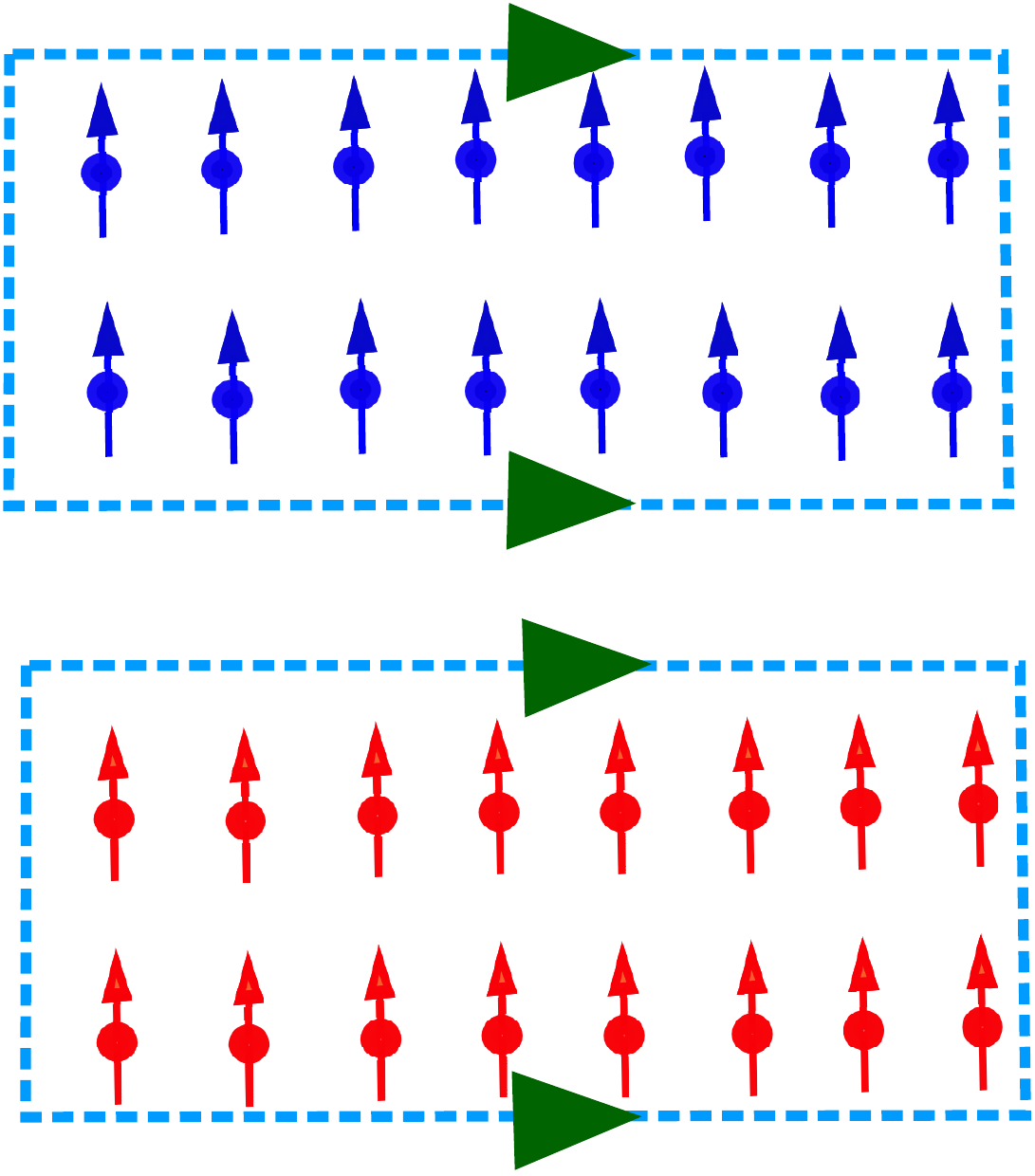}}
   \subfigure[\label{ee1}]{\includegraphics[width=.45\linewidth]{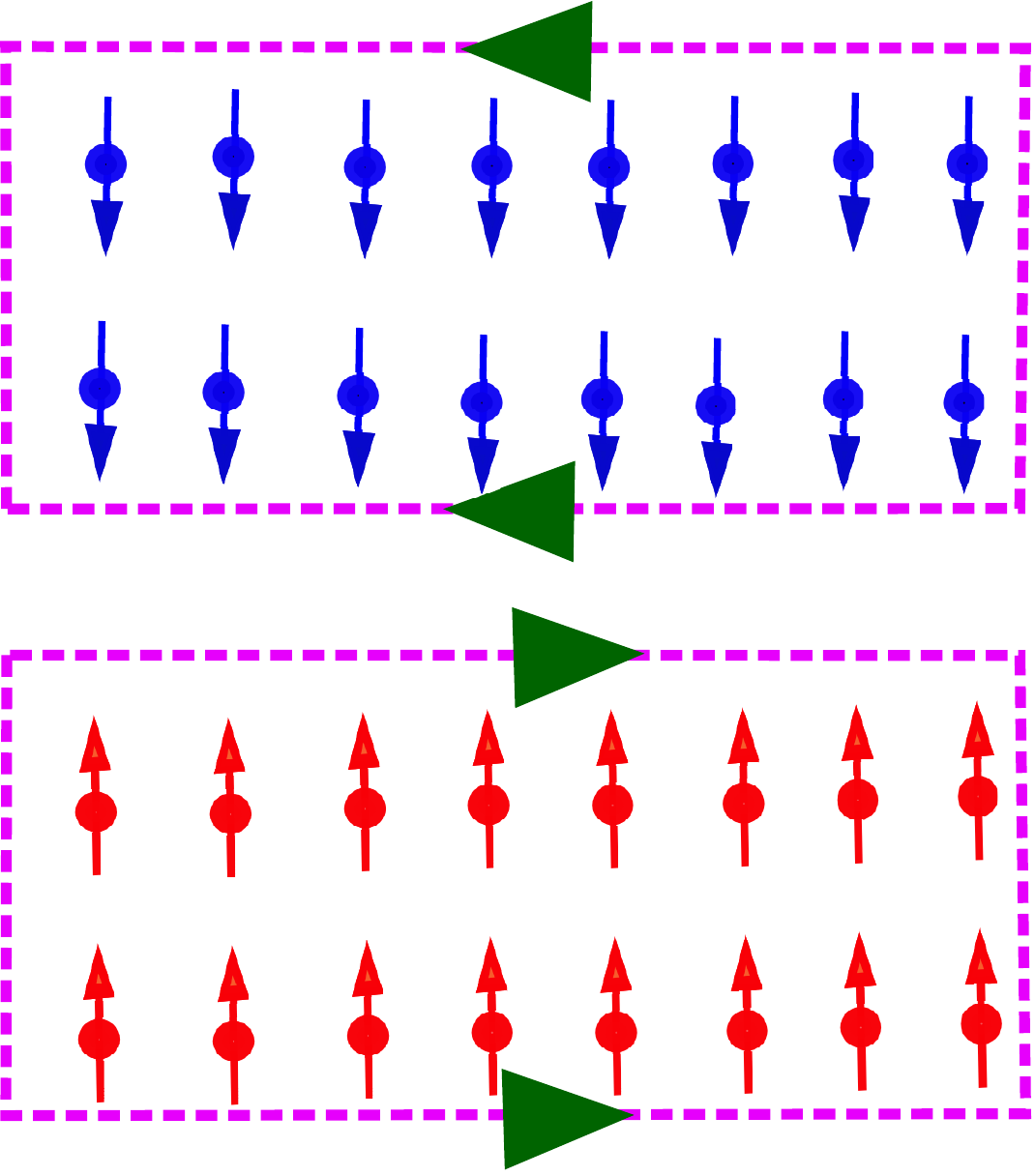}}
\caption{Color online. Schematics of chiral edge states (green arrows). $(a)$~Ferromagnetically coupled layers.  $(b)$ Antiferromagnetically coupled layers.  }
\end{figure}

\begin{align}
\mathcal A(\bold k)&=\epsilon_{0}\tau_0\otimes\sigma_0+ m_{\bo\phi}\tau_z\otimes\sigma_0, \\\mathcal B(\bold k)&=v_s\tau_0\otimes(f_\bo\sigma_+ + f_\bo^*\sigma_-)+|v_\perp|\tau_x\otimes\sigma_0,
\label{mata}
\end{align}
where $\epsilon_{0}=zv_s-z^\prime |v_s^\prime|+|v_\perp|+2v_t\sum_{\mu} \cos(\bo\cdot\bold{a}_\mu) \cos\phi$. As before $\mathcal A(\bold k)\neq \mathcal A^*(-\bold k)$, but $\mathcal B(\bold k)= \mathcal B^*(-\bold k)$. The Hamiltonian is diagonalized  as usual. The positive eigenvalues are given by
\begin{align}
\epsilon_{\alpha\pm}({\bf k})= \sqrt{m_{\bo\phi}^2+\epsilon_0^2-v_\perp^2-|v_s f_\bo|^2\pm 2g_\bo},
\end{align}
where $g_\bo=\sqrt{m_{\bo\phi}^2(\epsilon_0^2- |v_s f_\bo|^2) +|v_\perp v_s f_\bo|^2}$. 
The band structures depicted in Fig.~\ref{bandsA} are  very similar to that of bilayer ferromagnet with antiferromagnetic coupling for $h=0$.  As mentioned above, a finite magnetic field introduces spin canting. In this case, both the out-of-plane and in-plane DMIs contribute to the magnon excitations. This scenario is analyzed in the  Supplemental material: \href{http://stacks.iop.org/JPhysCM/28/47LT02/mmedia}{stacks.iop.org/JPhysCM/28/47LT02/mmedia}.
 
\begin{figure*}[ht]
\centering
  \subfigure[\label{e5}]{\includegraphics[width=.4\linewidth]{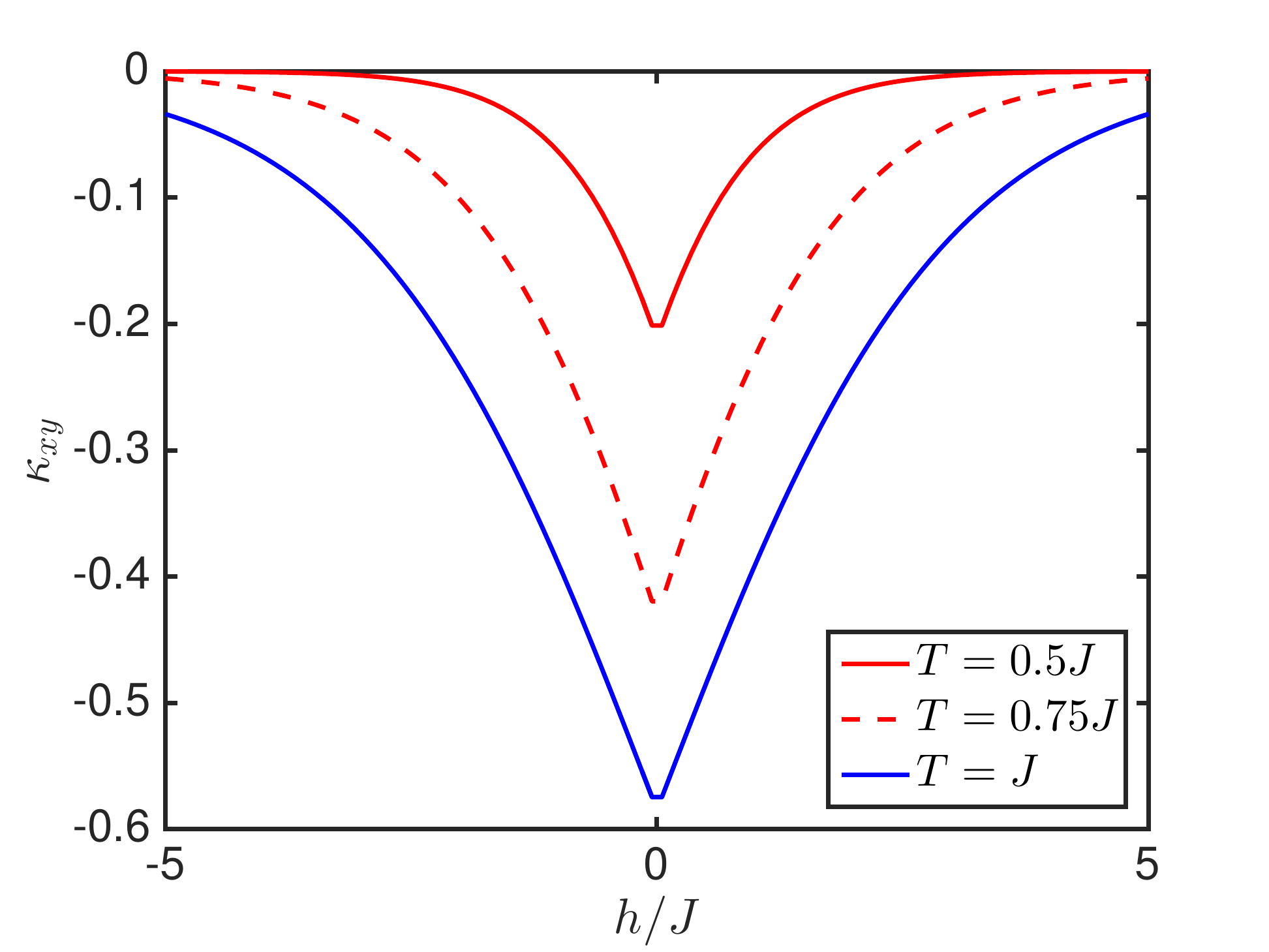}}
   \subfigure[\label{ee5}]{\includegraphics[width=.4\linewidth]{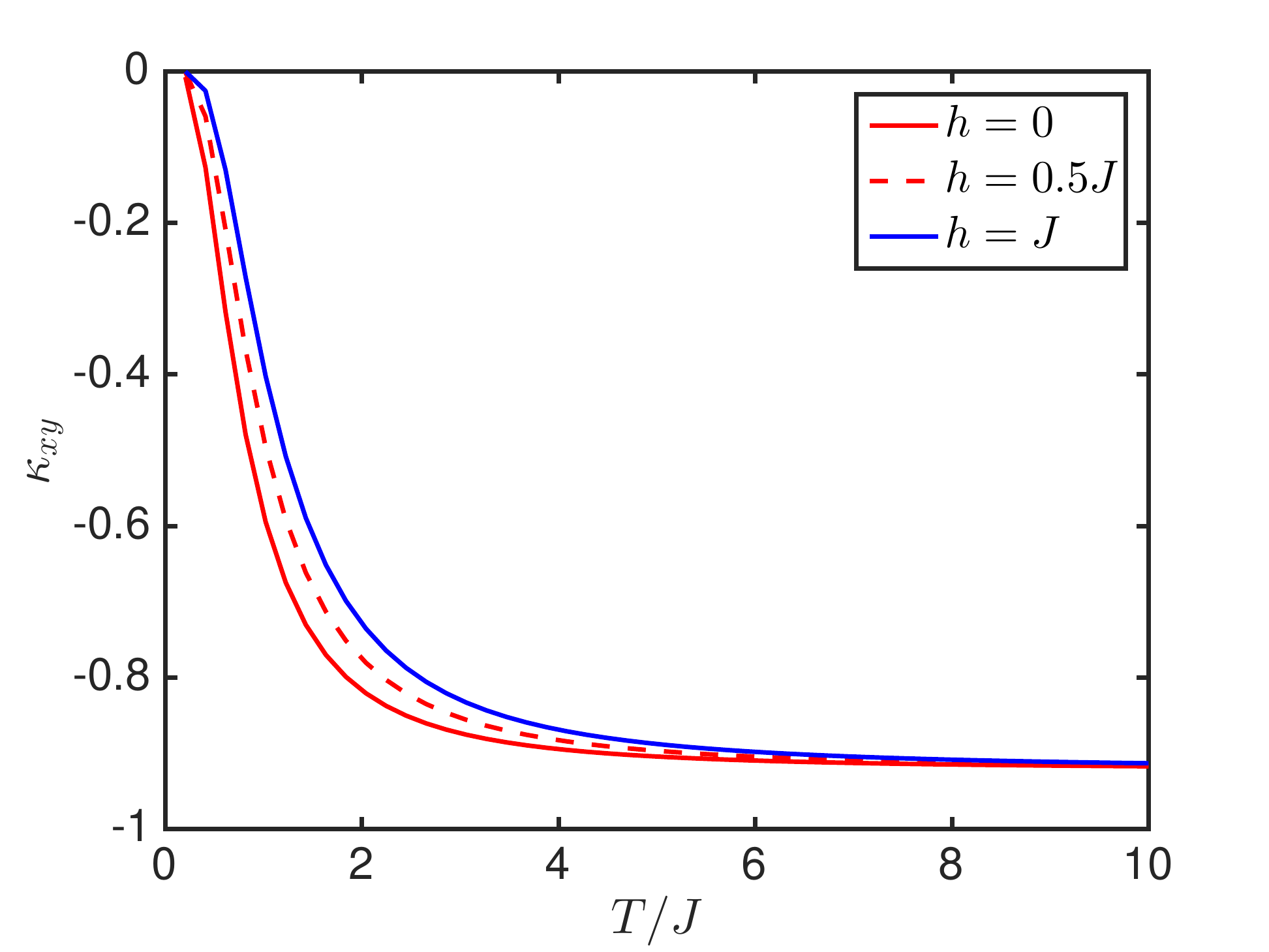}}
\caption{Color online. ~ Thermal Hall conductivity of ferromagnetically coupled AA-stacked bilayer honeycomb  chiral magnet as function of $(a)$ magnetic field  $(b)$~ temperature. The parameters are the same as Fig.~\ref{bands} (a). }
\label{thfm}
\end{figure*}

\begin{figure*}[ht]
\centering
  \subfigure[\label{e6}]{\includegraphics[width=.4\linewidth]{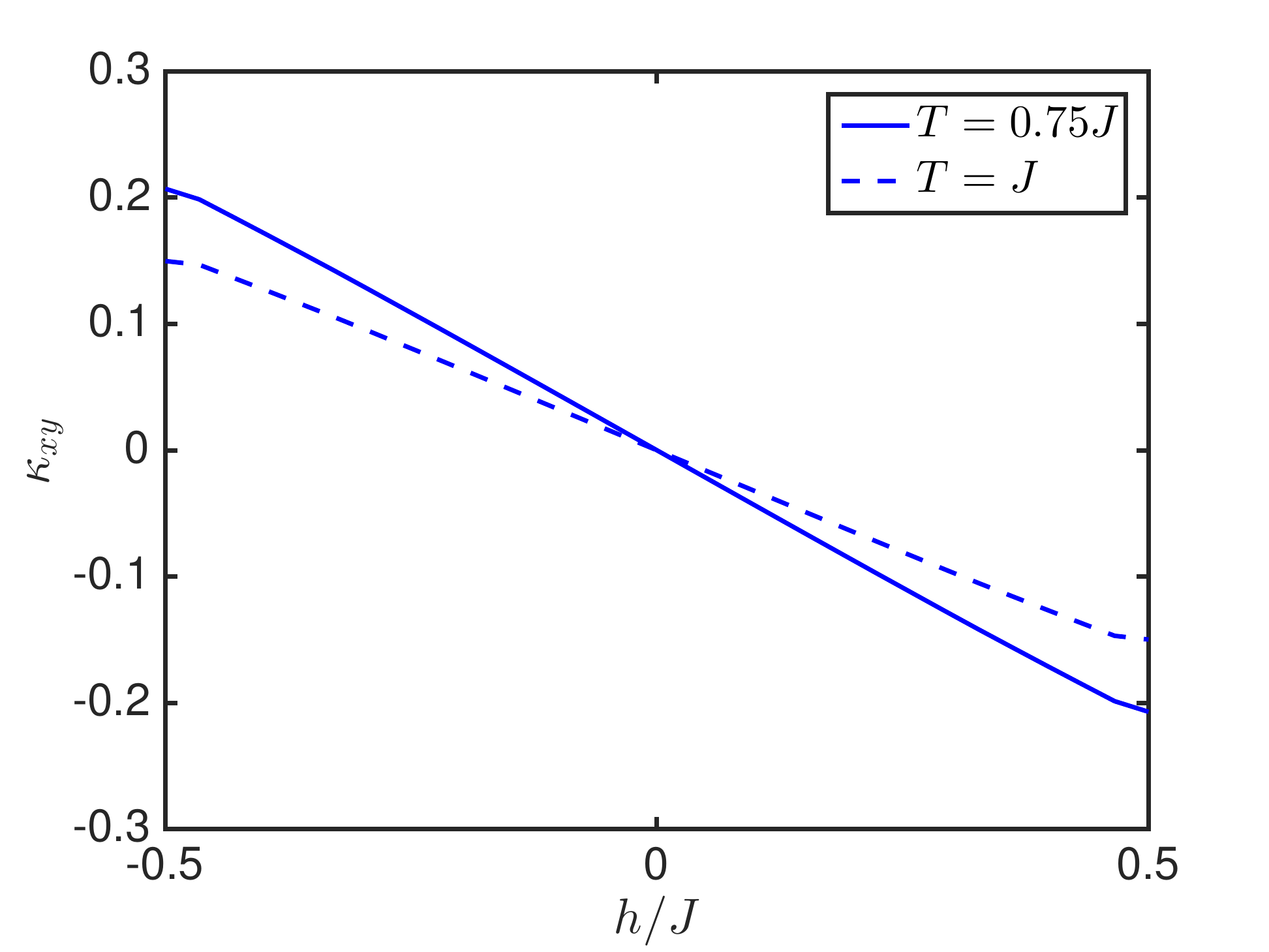}}
   \subfigure[\label{ee6}]{\includegraphics[width=.4\linewidth]{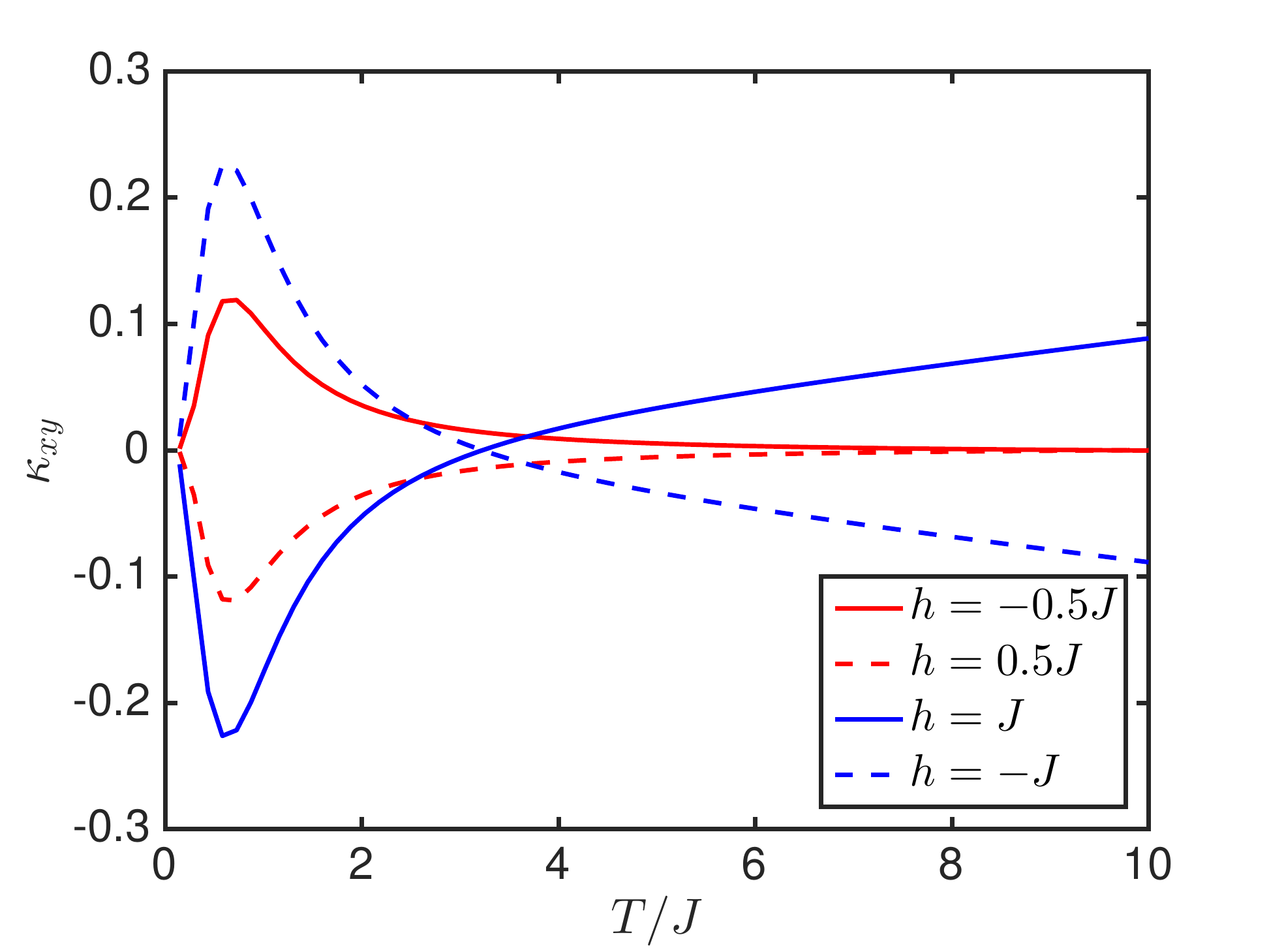}}
\caption{Color online. ~ Thermal Hall conductivity of antiferromagnetically coupled  AA-stacked bilayer honeycomb chiral magnet as function of $(a)$ magnetic field  $(b)$~ temperature. The parameters are the same as Fig.~\ref{bands} (b). }
\end{figure*}

\section{Magnon Transports}
\subsection{Magnon edges states}
Magnetic transports in topological magnon insulator materials are encoded in the protected chiral edge states of the system induced by the DMI.  Figure~\ref{e0} shows the evolution of the chiral protected edge states of the  ferromagnetically coupled layers in different parameter regimes. The chiral edge states  propagate in the same direction as depicted schematically in Fig.~\ref{e1}.  For antiferromagnetically coupled case, the same situation is observed with different parameters as depicted in Fig.~\ref{ee0}. However, the chiral  edge states propagate in opposite directions for the top and bottom layers because of opposite DMI as shown schematically in Fig.~\ref{ee1}.

\subsection{Magnon Hall effect}
The most interesting property of chiral  magnetic systems is the observation of magnon Hall effect \cite{alex1,alex1a, alex6, alex6a}. In magnon Hall effect \cite{alex0}, as well as magnon spin Nernst effect \cite{alex7}, the non-vanishing Berry curvatures induce an effective magnetic field in  the system, upon the application of a temperature gradient.    The propagation of magnons in the bilayer system is deflected by the chiral DMI.  Magnon Hall effect is characterized by a transverse thermal Hall conductivity,   given by \cite{alex2} $\kappa_{xy}=-{2k_B^2 T}V^{-1}\sum_{\bo\mu}c_2\lb n_\mu\rb\Omega_\mu(\bold k),$ where $V$ is the volume of the system, $k_B$ is the Boltzmann constant, $T$ is the temperature, $n_\mu\equiv n_B[\epsilon_\mu(\bold k)]=[e^{{\epsilon_\mu(\bold k)}/k_BT}-1]^{-1}$ is the Bose function,  $c_2(x)=(1+x)\lb \ln \frac{1+x}{x}\rb^2-(\ln x)^2-2\text{Li}_2(-x),$ and $\text{Li}_2(x)$ is a dilogarithm. Magnon spin Nernst conductivity has a similar definition \cite{alex7}   $\alpha_{xy}^s={k_B}V^{-1}\sum_{\bo\mu}c_1\lb n_\mu\rb\Omega_\mu(\bold k),$ where  $c_1(x)=(1+x)\ln(1+x)-x\ln x$. The chirality-induced Berry curvature is  defined as
\begin{align}
{\Omega}_\mu(\bold k)=-2\sum_{\mu\neq \mu^\prime}\frac{\text{Im}[ \braket{{\psi}_{\bo\mu}|v_x|{\psi}_{\bo\mu^\prime}}\braket{{\psi}_{\bo\mu^\prime}|v_y|{\psi}_{\bo\mu}}]}{[\epsilon_{\bo\mu}-\epsilon_{\bo\mu^\prime}]^2},
\label{chern2}
\end{align}
where $\ket{\psi_{\bo\mu}}$ are the eigenstates of the Hamiltonian and $\mu$ labels the bands; $v_{x,y}=\partial \mathcal{H}(\bold k)/\partial k_{x,y}$ defines the velocity operators.  

Figures~\ref{e5} and \ref{ee5} show the dependence of thermal Hall conductivity on the magnetic field  and the temperature for the ferromagnetically coupled layers. As the temperature approaches zero, $\kappa_{xy}$  vanishes due to lack of thermal excitations, but it never changes sign as the temperature increases or the magnetic field changes sign. This is what is observed theoretically in the single layer honeycomb chiral ferromagnet \cite{sol1}. However, for antiferromagnetically coupled layers shown in Figs.~\ref{e6} and \ref{ee6}, we see that $\kappa_{xy}$ changes sign as the magnetic field is reversed and vanishes at zero field. The sign change in $\kappa_{xy}$ is encoded in the magnon bulk bands, the Berry curvatures, and the propagation of the chiral edge states. The sign change in $\kappa_{xy}$ is very similar to what was observed on the pyrochlore chiral magnets upon reversing the direction of the applied magnetic field \cite{alex1, alex2}. Due to the Berry curvature, $\alpha_{xy}^s$ shows  similar trends (not shown). We also observe that for the chirality-proximity effect, where  only one layer contains a chiral DMI, topological effects are induced in the bilayer system and thermal conductivity $\kappa_{xy}$ is suppressed (see the Supplemental material: \href{http://stacks.iop.org/JPhysCM/28/47LT02/mmedia}{stacks.iop.org/JPhysCM/28/47LT02/mmedia}).

\section{Conclusion}

We have studied chirality-induced magnon transport in AA-stacked bilayer honeycomb chiral magnets. We observe remarkable distinctive features for   ferromagnetic and antiferromagnetic couplings.  In particular, the band structure and the chiral edge states  have different topological properties. As a result thermal Hall and spin Nernst conductivities  show a sign change for antiferromagnetic coupling in contrast to ferromagnetic coupling. As far as we know, chirality-induced transports and thermal Hall effect still await experimental observation on the honeycomb lattice. As mentioned  above, there are many accessible AA-stacked bilayer honeycomb quantum magnets in which chirality can be induced and these theoretical results can be confirmed. Experiments can also probe the observed magnon edge states, by noticing that spin-$1/2$ quantum magnets map to hardcore bosons. Thus, the  magnon edge states correspond to bosonic edge states, which can be studied experimentally in ultracold atoms on optical lattices similar to the realization of Haldane model \cite{jot}.  Our results can also be applied to  magnon spintronics  in chiral bilayer quantum magnetic systems.

\section*{Acknowledgments}  Research at Perimeter Institute is supported by the Government of Canada through Industry Canada and by the Province of Ontario through the Ministry of Research and Innovation.

\end{document}